\documentstyle[12pt,aaspp4,epsfig]{article}

\newcommand\bea {\begin{eqnarray}}
\newcommand\eea {\end{eqnarray}}
\newcommand\be {\begin{equation}}
\newcommand\ee {\end{equation}}
\newcommand\ie{{\it i.e.}}
\newcommand\Lag{{\cal L}}
\newcommand\calO{{\cal O}}
\newcommand\calM{{\cal M}}
\newcommand{\e}{{\mbox{e}}}
\newcommand\mN{m_N}

\newcommand\vk{{\vec k}}
\newcommand\vq{{\vec q}}
\newcommand\vp{{\vec p}}
\newcommand\vP{{\vec P}}
\newcommand\vbp{{\vec {\bar p}}}
\newcommand\vs{{\vec \sigma}}
\newcommand\vJ{{\vec J}}
\newcommand\hatr{{\hat r}}
\newcommand\mfm{{\mbox{fm}}}
\newcommand\MeV{{\mbox{MeV}}}
\newcommand\uC{u_{\rm C+N}}
\newcommand\vC{v_{\rm C+N}}

\begin{document}

\title{The Solar Proton Burning Process\\ 
           Revisited In Chiral Perturbation Theory}

\author{Tae-Sun Park\altaffilmark{1,2}, 
        Kuniharu Kubodera\altaffilmark{3}, 
        Dong-Pil Min\altaffilmark{2}
and Mannque Rho\altaffilmark{4,5}}
\altaffiltext{1}{Grupo de F\'\i sica Nuclear,
  Universidad de Salamanca, 37008 Salamanca, Spain \\
  e-mail: tpark@mozart.usal.es}
\altaffiltext{2}{Department of Physics and Center for Theoretical Physics,
  Seoul National University, Seoul 151-742, Korea \\
  e-mail(DPM): dpmin@phya.snu.ac.kr}
\altaffiltext{3}{Department of Physics and Astronomy, 
  University of South Carolina, Columbia, SC 29208, U.S.A. \\
  e-mail: kubodera@nuc003.psc.sc.edu}
\altaffiltext{4}{Service de Physique Th\'eorique, CEA Saclay, 
  91191 Gif-sur-Yvette Cedex, France \\
  e-mail: rho@spht.saclay.cea.fr}
\altaffiltext{5}{School of Physics,  Korea Institute for Advanced Study,
  Seoul 130-012, Korea}
\authoraddr{Tae-Sun Park, Room 37,
  F\'\i sica Te\'orica, Facultad de Ciencia, Edificio de Fisicas,
  Universidad de Salamanca, 37008 Salamanca, Spain,
  e-mail: tpark@mozart.usal.es}

\begin{abstract}
The proton burning process $p+p\rightarrow d +e^+ +\nu_e$,
important for the stellar evolution of main-sequence stars of mass
equal to or less than that of the Sun, is computed in effective
field theory using chiral perturbation expansion to the
next-to-next-to-leading chiral order. This represents a
model-independent calculation consistent with low-energy effective
theory of QCD comparable in accuracy to the radiative $np$ capture
at thermal energy previously calculated by first using very accurate
two-nucleon wavefunctions backed up  by an effective field theory
technique with a finite cut-off.
The result obtained thereby is found to support within  theoretical
uncertainties the previous calculation of the same process by
Bahcall and his co-workers.
\end{abstract}
\keywords{
nuclear reactions, nucleosynthesis, abundances
-- Sun: evolution
-- Sun: fundamental parameters
-- stars: evolution
-- stars: fundamental parameters}

\section{Introduction}

The proton fusion reaction
\be
p + p \rightarrow d + e^+  + \nu_e
\label{pp}\ee
which plays an important role for stellar evolution and -- as the
dominant neutrino source -- for the solar-neutrino problem, has
quite a long history of investigation. Indeed the reaction rate of
this process (hereafter called the $pp$ rate) was first calculated
by Bethe and Critchfield (\cite{bethe38}). Salpeter
(\cite{salpeter}) recalculated the $pp$ rate using the {\em effective
range approximation} and argued that the relevant nuclear matrix
element squared could be estimated with an accuracy of the $\sim$5
\% level. (The $pp$ rate itself was subject to much larger
uncertainty, $\sim$20 \%, because of the limited precision with
which the Fermi coupling constant was known at that time.) 
Bahcall and May (\cite{bahcall69})
examined the dependence of the $pp$
rate on explicit forms of the two-nucleon wavefunctions
generated by two-parameter nuclear potentials
of various forms
adjusted so as to reproduce the scattering length and effective
range (for the $pp$ channel) and the low-energy properties of the
deuteron (for the $np$ channel). The $pp$ rate was found to vary by
$\sim 1.5$ \% corresponding to the changes in the deuteron
wavefunction, and by $\sim 1.2$ \% due to the change in the $pp$
wavefunction. The most updated work along this line was done by
Kamionkowski and Bahcall (\cite{bahcall94}) employing deuteron
wavefunctions obtained from much more accurate potentials such as
the Argonne $v_{14}$, $v_{18}$, Urbana $v_{14}$, super-soft-core
(SSC) and Reid soft-core potentials. Changes in the $pp$ rates
arising from the different potentials were found to be $\sim 1\ \%$. 
Thus it seems that the presently available calculated $pp$
rate is robust and needs no further scrutiny, the famous solar
neutrino problem remaining unresolved from this angle and hence
persisting as one of the outstanding unsolved problems in
astrophysics (Bahcall \cite{unsolved}).

There are however two reasons for revisiting this issue. One is
that while the calculated $pp$ rate seems to have converged to a
``canonical"  value given in 
(Kamionkowski \& Bahcall \cite{bahcall94}, hereafter KB),
there lingers the
unsettling feeling that the strong interaction involved in nuclear
physics of the two-nucleon systems is infested with uncontrollable
uncertainties associated with model dependence in the treatment,
making it difficult to assess the accuracy achieved. Thus it is not
unexpected that this canonical value will be -- as has been in the past --
challenged. 
Indeed it has recently been argued by Ivanov et al. (\cite{ivanov})
that the nonrelativistic potential models used
in the previous works could be seriously in error. They show that
in their version of a {\it relativistic field theory model}, the
$pp$ rate comes out to be as big as $2.9$ times the previous
estimates.\footnote{They use  a procedure that seems to disagree with other
physical properties of low-energy $pp$ systems, as was pointed out by
Bahcall and Kamionkowski (1997). It has also been pointed out by
Degl'Innocenti et al. (\cite{Degl})
that such a large deviation from the value
used by KB would be inconsistent with helioseismology in the Sun.}
Should their new result turn out to be
correct, it would have profound consequences on theories of
stellar evolution in general and on the solar neutrino problem in
particular. In a nutshell, the issue comes down to whether or not
a more general framework such as relativistic field theory would
invalidate the calculation made in the traditional nonrelativistic
potential models. 
The claim of the authors in (Ivanov et al. \cite{ivanov}) is that 
it indeed does. Our aim is to address this issue 
using a low-energy effective field theory of QCD that has found a 
quantitative success in other nuclear processes.

The second reason is really more theoretical, independent of the
above important astrophysical issue. Along with the thermal $np$
capture, the proton fusion process is the simplest nuclear process
amenable to an accurate calculation -- something rare in hadronic
physics -- and it is of interest on its own to test how well a
calculation faithful to a ``first-principle approach" can tackle
this problem. In particular, we are interested in checking how
accurately the effective field theory approach, found to be
stunningly successful for the $np$ capture $n+p\rightarrow
d+\gamma$, low-energy NN scattering and static properties of the
deuteron (Park, Min, \& Rho \cite{pmr_PRL};
Park et al. \cite{pkmr}), 
fares with the proton fusion problem,
the weak interaction sector of the Standard Model.

The strategy we shall adopt here is quite close to that used for the
$np$ capture process (Park et al. \cite{pmr_PRL}). We shall use chiral
perturbation theory to the next-to-next-to-leading order (NNLO) 
in chiral counting; as defined precisely below, 
this corresponds to $O(Q^3)$ relative to the leading-order term.
This is roughly the same order as considered for the $np$ capture. 
However the relative importance of terms of various chiral orders
is somewhat different here.  As we explain later, 
in the present case, the corrections to the leading order 
are not suppressed by what is called 
the ``chiral filter" \footnote{The chiral filter phenomenon is
explained in detail in (Park, Min, \& Rho, \cite{pmr_Report}). 
Crudely stated, it refers to the general feature 
that whenever one soft-pion exchange is allowed by kinematics and
selection rules, it should give a dominant contribution
with higher-order (or shorter-range) terms strongly suppressed.  
A corollary to this is that whenever
one soft-pion exchange is not allowed, 
all higher-order terms {\it can be} important, 
making chiral perturbation calculation generically less powerful.} 
and so the accuracy with which
these can be calculated is not as good as in the $np$ capture case.
Even so, using the argument developed in (Park et al. \cite{pkmr}), 
we shall suggest that
the procedure used here provides a model-independent result in the
same sense as in (Park et al. \cite{pmr_PRL}, Park et al. \cite{pkmr}).

To streamline the presentation, we first give our result 
and then discuss (as briefly as possible) 
how we arrive at it in the rest of the paper. 
Apart from the meson-exchange contributions which are of order of
${\cal O}(Q^3)$ and which for the reason mentioned above and further 
stressed in our concluding section, are the main uncertainty to the order
considered,
our chiral perturbation theory result in terms of the
reduced matrix element $\Lambda$
defined in (Bahcall \& May \cite{bahcall69}) is
\bea
\Lambda_{\chi PT}^2
= (1 \pm 0.003) \times 6.93
\label{ChPT}
\eea
where the uncertainty is due to experimental errors.\footnote{
Our theoretical uncertainty is, if very conservatively estimated,
about $0.1\ \%$.}
The above result is to be compared 
with the value obtained by Kamionkowski and Bahcall (KB) 
\be
\Lambda_{\rm KB}^2 = 
 \left(1 \pm 0.002^{+0.014}_{-0.009}\right) \times 6.92\, .
\label{bahcallS}\ee
As we shall explain, there are some differences in details
between our calculational framework and that of KB,
but our final numerical result is in good agreement 
with that of KB
and disagrees with that of Ivanov et al. (\cite{ivanov}).

The paper is organized as follows. 
In Section \ref{2}, our strategy for carrying out a
chiral perturbation calculation for two-nucleon systems is outlined. 
Our approach here is similar to the one used 
in the previous calculation of the $np$ capture process. 
We shall sketch a justification of this approach 
from the standpoint of low-energy effective field theory 
of QCD (with concrete supporting evidence 
summarized in Subsection \ref{cutoff}).
Section \ref{chiralcounting} describes chiral counting of the terms 
appearing in the relevant weak current. 
In Section \ref{wavefunction}, the wavefunctions 
for the initial $pp$ state and the final $d$ state are specified.
Our numerical results are given in Section \ref{numbers}, 
and a brief discussion including a comment on the main uncertainty 
in the calculation is given in Section \ref{discussion}.

\section{Effective Field Theory for Nuclei}\label{2}

Before presenting our calculation, we sketch the chain of
reasoning that leads us to believe that the result we obtained is
model-independent. The argument is essentially identical 
to that used for thermal $np$ capture, 
and therefore we review briefly the key ingredients of
the calculations in (Park et al. \cite{pmr_PRL}; Park et al. \cite{pkmr})
for the process
\be
n+p\rightarrow d+\gamma.\label{np}
\ee

The basic premise of our approach is an application of the
``Weinberg theorem" (see, e.g., Weinberg \cite{theorem}), 
which for the case in hand
implies that at very low energies 
at which the process (\ref{np}) is
probed, QCD can be described by chiral perturbation theory. 
In setting up chiral expansion for two-nucleon systems 
we are interested in, we follow Weinberg (\cite{weinberg}) 
and separate the relevant transition amplitude into
the ``reducible" and ``irreducible" terms. Bound states and
resonances involve reducible graphs 
that are kinematically enhanced. 
To incorporate this feature,
we need to sum up a given set of reducible terms 
to all orders, a procedure equivalent 
to solving the Schr\"odinger or Lippmann-Schwinger equation. 
The irreducible graphs that appear as kernels in the
latter are calculated to a specified order in chiral expansion.
How this can be done in a systematic way consistent with chiral
counting has been extensively discussed in the literature
(Ordonez, Ray, \&  van Kolck \cite{vankolck};
Kaplan, Savage, \& Wise \cite{ksw};
Luke \& Manohar \cite{lm}; 
Beane, Cohen, \& Phillips \cite{cohen}).
In (Park et al. \cite{pmr_PRL}), we have written
the relevant M1 matrix element for the process (\ref{np}) with the
EM current given by the irreducible graphs calculated to NNLO and
accurate wavefunctions for the initial and final states, namely
the Argonne $v_{18}$ wave function. 
The Argonne $v_{18}$ potential 
(Wiringa, Stoks, \& Schiavilla \cite{v18}) 
provides an impressive description of {\it all}
two-nucleon data at low energies up to $\sim$ 350 MeV. 
So the wavefunctions obtained with it correspond to summing both the
irreducible and reducible graphs to {\it all} orders. 
It seems then that the procedure used in (Park et al. \cite{pmr_PRL}) 
is not quite consistent
since the current is computed only to NNLO. 
We argued however in (Park et al. \cite{pmr_PRL}) 
that this is consistent with {\it the chiral counting 
to the chiral order considered} and the error made in the
terms of higher order than that taken in the current cannot be
important. 
We have subsequently demonstrated (Park et al. \cite{pkmr}) 
the correctness of this conjecture 
by showing in a cut-off field theory
that a systematic summation 
of the reducible diagrams with the irreducible
graphs appearing as a kernel -- which accurately describes all
static properties of the deuteron and NN scattering lengths -- 
gives precisely the same result 
for the leading chiral order M1 matrix element 
as that obtained with the Argonne $v_{18}$ wave function. 
Now, as we know, the ratios of subleading chiral terms 
to the leading term are almost model-independent and
insensitive to the renormalization schemes used, 
so these {\it ratios} can be calculated with reasonable accuracy. 
Thus the result obtained in (Park et al. \cite{pmr_PRL}),           
\be
\sigma^{\chi PT}=334\pm 3\ \ {\rm mb}\label{npth}
\ee
(where the error represents our ignorance regarding 
the short-distance part of the interaction 
in chiral perturbation theory) which agrees well with the experiment,
\be
\sigma^{exp}=334.2\pm 0.5\ \ {\rm mb},
\ee
can be taken as a first-principle calculation as argued in
(Park et al. \cite{pkmr}).

For the weak process (\ref{pp}) we will follow the same procedure 
and, to supply support for it, carry out 
an effective field theory calculation up to NNLO
with the use of the same cut-off scheme as in (Park et al. \cite{pkmr}),
where the leading-order M1 matrix element was evaluated. 

\section{Chiral Counting and the Weak Current}\label{chiralcounting}

As in KB (\cite{bahcall94}), we shall include 
the vacuum polarization effect 
pertinent to the initial $pp$ channel
as well as the effect of meson-exchange currents.
In addition, we include the effect of
the two-photon exchange Coulomb interaction for the $pp$ channel.
\subsection{Chiral counting}
\indent

The core temperature of the Sun is believed to be about 
$T_C = 15.5\times 10^6\ K$, for which the
corresponding nucleon momentum is 
$p \sim \sqrt{2 m_p T_C}\sim 1.6$ MeV.
We are therefore interested in the process (\ref{pp})
at very low energies, a situation similar to 
the thermal $np$ capture process. This is an arena where
the low-energy effective field theory can be powerful. 
At these low energies, the reaction is dominated 
by the $^1S_0\rightarrow d$ transition, 
involving only the transition operators that carry 
orbital angular momentum $L=(0,\,2)$, spin $S=1$ 
and total angular momentum $J=1$. 
The momentum carried  by the leptons,
$\vk=\vp_e+\vp_\nu$, is also very small, $ 0 \le |\vk| \le
\left[4 m_d m_p - 2 m_d \left(
  4 m_d m_p + m_e^2 - m_d^2\right)^{\frac12}\right]^{\frac12}
\simeq 0.78\ \ \MeV$,
where $m_p$, $m_d$ and $m_e$ are the mass of proton, deuteron
and electron, respectively.
We can therefore safely ignore terms of ${\cal O}(|\vk|^2)$ and
work only to ${\cal O}(|\vk|)$. These selection rules will be
implicit throughout our subsequent arguments

To carry out the procedure explained in Section \ref{2}, 
we need to specify a counting rule for irreducible terms 
for the electroweak hadronic current 
$J^{\mu,-} \equiv J^{\mu,1} - i J^{\mu,2}$, where
\be
J^{\mu,a} = V^{\mu,a} - A^{\mu,a}.
\ee
This rule has been extensively discussed in
(Park et al. \cite{pmr_PRL}; Park et al. \cite{pmr_Report};
Park, Jung, \& Min \cite{pjm_gA}), 
so we shall simply summarize it here.

Let $Q$ denote the typical momentum scale probed in the process and
$\Lambda_\chi$ be the chiral scale $\sim m_N$. The chiral expansion
is made in powers of $Q/\Lambda_\chi$.
A Feynman diagram for an $A$-nucleon process 
(Weinberg \cite{weinberg}) goes
as $Q^\nu$.
Now  for a Feynman graph with $L$ loops, $N_E$
external fields and $C$ separately connected pieces, the exponent
$\nu$ is given by
\begin{equation}
\nu = 4-A-2C+2L- N_E +\sum_{i}\nu_i\;,
\;\;\;{\rm with}\; \nu_i=d_i+ e_i +{n_i\over 2}-2,
\end{equation}
where $n_i$ is the number of nucleon lines, $d_i$ the number of
derivatives or powers of $m_{\pi}$ and $e_i$ the number of external
fields at the $i$-th vertex.  Since $\nu_i$ is defined so that chiral
symmetry guarantees $\nu_i\ge 0$ (Weinberg \cite{weinberg}) even in the
presence of external fields (Rho \cite{rho2}), the effective
Lagrangian can be ordered according to $\nu_i$:
\begin{equation}
{\cal L}_{eff} = {\cal L}_0 + {\cal L}_1 +{\cal L}_2 +\cdots,
\end{equation}
where ${\cal L}_n$ has vertices with $\nu_i=n$.

\subsection{Current}
\indent

Let us now apply this counting rule to our case. The leading
contribution arises from a one-body diagram with 
$\nu= \nu_0 \equiv -3$ (\ie, $L=0$, $C=A=2$, $N_E=1$ and $\nu_i=0$),
and the next-to-leading order ($\nu=\nu_0+1$) 
also comes from a one-body diagram involving a $\nu_i=1$ vertex. 
These one-body operators are given by 
\bea
V_{\rm 1B}^{\mu,a} &=&
\sum_l \frac{\tau^a_l}{2}
 \left( 1, \ \frac{\vp_l + \vp_l^{\,\prime}}{2 \mN}
  + \frac{i \mu_V}{2 \mN} \vk\times\vs_l
 \right) ,
\\
A_{\rm 1B}^{\mu,a} &=&
g_A \sum_l \frac{\tau^a_l}{2} \left(
   \frac{\vs_l\cdot(\vp_l+ \vp_l^{\,\prime})}{2\mN},\ \vs_l 
\right)\,,
\eea
with $\vp_l$ ($\vp_l^{\,\prime}$) the incoming (outgoing) momentum of the
$l$-th nucleon, $g_A\simeq 1.2601(25)$ 
(Particle Data Group \cite{PDG96}), 
$\mu_V \equiv \mu_p - \mu_n \simeq 4.7059$ 
the nucleon isovector magnetic moment and $m_N$ the nucleon mass.
Of these one-body operators only the Gamow-Teller term 
contributes at the leading order, while the weak magnetism (WM) term 
that comes at the next order is further suppressed
kinematically for the small lepton momentum $|\vk| \ll Q$. 
The others do not contribute by the selection rule. 
We will keep the WM term but neglect all other higher
chiral order $\vk$-dependent terms. 
Thus the one-body current we calculate is
\bea
\vJ_{\rm 1B}^- &=& \vJ_{\rm GT}^- + \vJ_{\rm WM}^- + \cdots,\label{1-bod}
\label{1B}
\eea
with
\bea
\vJ_{\rm GT}^- &=& - g_A \sum_l  \tau_l^- \vs_l ,
\label{JGT} \\
\vJ_{\rm WM}^- &=& \frac{i \mu_V}{2 \mN} \vk\times \sum_l  \tau_l^- \vs_l.
\label{JMM}
\eea
To proceed to higher orders, it is convenient to do chiral
counting relative to the leading order. 
Expressed in this way, the leading order one-body current 
eq.(\ref{1-bod}) is taken to be $\sim \calO(1)$, 
and the next-to-leading order term is $\sim \calO(Q)$. 
The two-body exchange current can enter at the next-to-next-to-leading
order (NNLO), that is at $\calO(Q^2)$. 
Because of the selection rule, however, 
there is no term of this order contributing to the process (\ref{pp}). 
Hence the two-body exchange current at tree order enters at $\calO(Q^3)$. 
Note the difference in the roles of the exchange currents
between the present case and the $np$ capture process. 
In the latter, the exchange-current (2-body) M1 operator 
is $\sim \calO(Q)$ relative to the leading one-body M1 operator. 
This is due to the fact that the one-body M1 operator is 
formally suppressed by one order of chiral counting, 
so that the one-soft-pion exchange two-body current is enhanced
relative to the one-body operator. This is just the chiral
filter enhancement referred to above. 
In the Gamow-Teller case, the chiral filter
mechanism is not operative, making the leading two-body correction
come at $\calO(Q^3)$. This renders evaluation of the two-body terms
less precise although their overall importance relative 
to the one-body term should be considered diminished here.
We will return to this matter later.

Calculation of the meson exchange Gamow-Teller operator is
discussed in (Park et al. \cite{pjm_gA}). Here we skip the details and
simply quote the result. 
The leading two-body contributions of $\calO(Q^3)$ come from the
one-pion exchange and four-Fermi contact graphs with $L=0$,
$C=A-1=1$ and one $\nu_i=1$ interaction ($\nu=\nu_0+3$). The
relevant part of the Lagrangian with the
$\nu_i=1$ interaction is of the form 
(Bernard, Kaiser, \& Meissner \cite{meissner}; Cohen et al. \cite{Kolck3})
\bea
\Lag_1 &=& {\bar B} \left(
  \frac{v^\mu v^\nu - g^{\mu\nu}}{2 m_N} D_\mu D_\nu
   + 4 c_3 i\Delta\cdot i\Delta
   + \left(2 c_4 + \frac{1}{2m_N}\right)
  \left[S^\mu, \,S^\nu\right] \left[ i \Delta_\mu,\, i\Delta_\nu\right]
\right)B
\nonumber \\
 && -\ 4 i d_1 \,
 {\bar B} S\cdot \Delta B\, {\bar B} B
 + 2 i d_2 \,
  \epsilon^{abc}\,\epsilon_{\mu\nu\lambda\delta} v^\mu \Delta^{\nu,a}
 {\bar B} S^\lambda \tau^b B\, {\bar B} S^\delta \tau^c B
+ \cdots,
\label{Lag1}\eea
where $\epsilon_{0123}= +1$, and the definitions of the 
``covariant derivatives"
$D_\mu$, $\Delta_\mu$ and the spin operator $S^\mu$ can be found
in (Bernard et al. \cite{meissner}; Cohen et al. \cite{Kolck3}).
There are no vector-current contributions and the surviving
two-body
Gamow-Teller operator in momentum space is
\bea
{\vec A}_{{\rm 2B}}^a &=&
- \frac{g_A}{2 \mN f_\pi^2}\, \frac{1}{m_\pi^2 + \vq^2}
 \left\{ - \frac{i}{2} \tau_\otimes^a\,\vp\,\vs_\ominus\cdot\vq
 \right.
 \nonumber \\
 &&\left. \ \ \ +\
  \hat c_3 \vq \, \vq\cdot
  (\tau_\oplus^a \vs_\oplus + \tau_\ominus^a \vs_\ominus)
  + \left(\hat c_4 + \frac14\right) \tau_\otimes^a\,
 \vq \times (\vs_\otimes\times \vq)
 \frac{}{}\right\}
\nonumber \\
 &-& \frac{g_A}{2 m_N f_\pi^2} \left[
  \hat d_1 (\tau_\oplus^a \vs_\oplus + \tau_\ominus^a \vs_\ominus)
 - 2 \hat d_2 \tau_\otimes^a \vs_\otimes
 \right]
\eea
with $\vp \equiv \frac12 (\vbp_1 - \vbp_2)$, 
$\vbp_l\equiv \frac12(\vp_l+ \vp_l^{\,\prime})$;
$f_\pi\simeq 93\ \MeV$ is the pion decay constant
and $m_\pi$ is the pion mass.
We have defined $\vs_{\oplus, \ominus, \otimes}$ by
\be
\vs_{\oplus, \ominus} \equiv \vs_1 \pm \vs_2, \ \ \
\vs_\otimes \equiv \vs_1 \times \vs_2,
\ee
and similarly for $\tau_{\oplus,\ominus, \otimes}^a$.
The dimensionless parameters ${\hat c}_i$ and ${\hat d}_i$ are defined as
\be
c_i \equiv \frac{1}{m_N} \hat c_i \ \ \
\mbox{and}\ \ \
d_i \equiv \frac{g_A}{m_N f_\pi^2} \hat d_i\,.
\label{c-normal}\ee
Among the $1/\mN$ corrections,  terms proportional to
$\vP \equiv \vbp_1 + \vbp_2 = - \frac12 \vk$ have been dropped.
The $\vp$-dependent terms, which come from $\frac{1}{2 m_N} {\bar B}
{\vec D}^2 B$ in eq.(\ref{Lag1}), shall be referred to as
the ``kinetic term". 
The low-energy constants $c_3$ and $c_4$ have been
determined from pion-nucleon experiments by Bernard et al. (\cite{meissner}):
\bea
c_3 &=& -5.29 \pm 0.25\ \mbox{GeV}^{-1},
\nonumber \\
c_4 &=&  3.63 \pm 0.10\ \mbox{GeV}^{-1}.
\label{c3c4}\eea
They are almost completely saturated by the
resonance-exchange contributions
\bea
c_3^{\rm Res} &=& c_3^\Delta + c_3^S + c_3^R
  = \left(-3.83 -1.40 - 0.06\right) \ \mbox{GeV}^{-1}
  = -5.29 \ \mbox{GeV}^{-1} , \nonumber \\
c_4^{\rm Res} &=& c_4^\Delta + c_4^\rho + c_4^R
  = \left(1.92 + 1.63 + 0.12\right) \ \mbox{GeV}^{-1}
  = 3.67 \ \mbox{GeV}^{-1}
\eea
where the superscripts $\Delta$, $\rho$, $S$  and  $R$
denote the contributions from the exchange of
$\Delta(1232)$, $\rho$-mesons, scalar-mesons
and the Roper resonance, respectively.
In terms of the dimensionless constants $\hat c_3$ and $\hat c_4$,
 eq.(\ref{c-normal}) gives us
\bea
\hat c_3 &\simeq& - 4.97 \pm 0.23,
\nonumber \\
\hat c_4 &\simeq& 3.41 \pm 0.09 .
\label{hat_c}\eea
There are no low-energy data to determine the constants 
$\hat d_1$ and $\hat d_2$. However, it was argued in 
(Park et al. \cite{pmr_PRL})
that the zero-range operators corresponding to 
the $\hat d_1$ and $\hat d_2$ terms should be killed 
by the short-range correlation. 
We shall justify this argument here by showing that for ``natural"
coefficients in the Lagrangian (as required by chiral symmetry)
their matrix elements are indeed negligible.

Finally, when Fourier-transformed, 
the two-body  current in coordinate space reads
\bea
{\tilde J}_{\rm 2B}^{i,a}(\vec r) &=&
- {\tilde A}_{{\rm 2B}}^{i,a}(\vec r) + \cdots
\nonumber \\
 &=&
 - \frac{g_A m_\pi^3}{2 \mN f_\pi^2}\,
 \left\{ \frac{y_1(m_\pi r)}{2 m_\pi^2 r}
   \tau_\otimes^a\,p^i\,\vs_\ominus\cdot\hatr
 \right.
 \nonumber \\
 &&\left. \ \ \ +\
  \hat c_3 \left(\hatr^{ij} y_2(m_\pi r) + \frac{\delta^{ij}}{3}
     {y}_0(m_\pi r)\right)
  (\tau_\oplus^a \vs_\oplus^j + \tau_\ominus^a \vs_\ominus^j)
  \right.
 \nonumber \\
 &&\left. \ \ \ +\
  \left(\hat c_4 + \frac14\right) \tau_\otimes^a\,
   \left(- \hatr^{ij} y_2(m_\pi r) + \frac{2 \delta^{ij}}{3}
     {y}_0(m_\pi r)\right) \vs_\otimes^j
 \frac{}{}\right\}
\nonumber \\
 &+& \frac{g_A}{2 m_N f_\pi^2} \left[
  \left(\hat d_1 + \frac13 \hat c_3\right)
    (\tau_\oplus^a \vs_\oplus^i + \tau_\ominus^a \vs_\ominus^i)
  + \left(- 2 \hat d_2 + \frac23 \hat c_4 + \frac16\right)
       \tau_\otimes^a \vs_\otimes^i
 \right] \delta^{(3)}(\vec r)
\nonumber \\
&+& \cdots
\label{vJ2B}\eea
where the ellipses denote irrelevant terms
and the Yukawa functions, $y$'s, are  defined as
\bea
y_0(x) &=& \frac{\e^{-x}}{4\pi x},
\nonumber \\
y_1(x) &=& \left(1+x\right) \frac{\e^{-x}}{4\pi x}
 = - x \frac{d}{dx} y_0(x),
\nonumber \\
y_2(x) &=& \left(1+\frac3x+\frac3{x^2}\right) \frac{\e^{-x}}{4\pi x}
 = x \frac{d}{dx}  \frac1x \frac{d}{dx} y_0(x).
\eea

\section{Wavefunctions}\label{wavefunction}

In the scheme we are using, it is essential to have accurate
wavefunctions for the initial and final states so that the
leading single-particle Gamow-Teller matrix element is accurately
given. In (Park et al. \cite{pmr_PRL}), 
we have shown that the single-particle M1
matrix element for the $np$ capture process (\ref{np}) is
accurately given by the wavefunctions obtained with the Argonne
$v_{18}$ potential. The single-particle Gamow-Teller operator is
identical in structure to the isovector M1 operator, so provided
that electromagnetic effects are suitably taken into account, we
expect that the Gamow-Teller matrix element for the proton fusion
process (\ref{pp}) will also be given very accurately by the
wavefunctions of the Argonne $v_{18}$ potential. Once this argument
is justified, then we can conclude immediately that the answer was
{\it essentially} given by Bahcall and collaborators since they
used, among others, the Argonne $v_{18}$ potential in calculating
the Gamow-Teller matrix element. One difference in their
calculation from ours is that in their treatment, the initial and
final wavefunctions were calculated asymmetrically with the initial
scattering wavefunction treated approximately while the final
deuteron state treated ``exactly". Our strategy in the chiral
perturbation approach requires that they be treated with the same
``exact" potential. Numerically, however, their approximation for
the initial state should be quite good, so it is not unreasonable
to expect that not much will be gained by the symmetric treatment.
We confirm this.

We now turn to discussion of the wavefunctions used in our calculation.

\subsection{The deuteron wavefunction}
\indent

The deuteron wavefunction obtained with the Argonne $v_{18}$
potential (referred to as the ``Argonne potential" in what follows) is
well-known, so without going into any details, we will merely
define the notations for later use.

The deuteron wavefunction will be written in the conventional form
\be
\psi_{d}(\vec r) = \frac{A_S}{\sqrt{4\pi}r}
\left[u_d(r) + \frac{S_{12}({\hat r})}{\sqrt{8}} w_d(r)\right]
 |^3S_1\rangle ,
\ee
where
$S_{12}({\hat r})\equiv 3 \vs_1 \cdot \hatr \,\vs_2\cdot \hatr
 - \vs_1\cdot \vs_2$.
Asymptotically, the radial functions take the form
\bea
\lim_{r\rightarrow \infty}u_d(r) &=& \phi_d(r) \equiv
 \e^{-\gamma r},
\nonumber\\
\lim_{r\rightarrow \infty}w_d(r) &=& \eta_d \left(1+ \frac{3}{\gamma r}
 + \frac{3}{(\gamma r)^2} \right) \phi_d(r)
\eea
where $\gamma$ is the wave-number,
$\gamma = \sqrt{m_N B_d}$, with $B_d$ the binding energy;
$\eta_d$ is the deuteron $D/S$ ratio,
and the normalization factor $A_S$ is defined by
\be
A_S^2 \int_0^\infty\! dr\,\left[u_d^2(r) + w_d^2(r)\right]=1.
\label{Asdef}\ee
The experimental values 
(van der Leun \& Alderliesten \cite{BdExp}; 
Ericson \&  Rosa-Clot \cite{ASExp}; 
Rodning \& Knutson \cite{etaExp}) are 
\be
B_d^{\rm exp} = 2.224575(9)\ \MeV, \ \
A_S^{\rm exp} = 0.8846(8)\ \mfm^{\frac12}, \ \
\eta_d^{\rm exp} = 0.0256(4).
\label{d:exp}\ee
It is also interesting to note that
$A_S$ can be expressed in terms of the
effective range defined at the pole-position,
\bea
\rho_d &\equiv&
 2 \int_0^\infty\! dr\,\left[\phi_d^2(r) -
\frac{u_d^2(r) + w_d^2(r)}{1+ \eta_d^2}\right]
\,=\,
 \frac{1}{\gamma} - \frac{2}{(1+ \eta_d^2) A_S^2}
\label{rhoddef}\eea
or
\be
A_S^2 = \frac{2 \gamma}{(1 + \eta_d^2) (1 - \gamma \rho_d)}.
\label{AsER}\ee
With eq.(\ref{rhoddef}), the experimental data eq.(\ref{d:exp}) lead to
\be
\rho_d^{\exp} = 1.7635(46)\ \mfm .
\label{rdexp}\ee

One can get an idea as to
how accurately the Argonne wavefunction describes the
properties of the deuteron by looking at the predictions:
$B_d = 2.22460\ \MeV$,
$A_S = 0.88506\ \mfm^{\frac12}$,
$\eta_d = 0.02504$
and $\rho_d^{v18} = 1.7660\ \mfm$.
The close agreement between the Argonne results and the experimental values
invites us to use the two interchangeably in referring to ``experiments".
Thus for instance in (Park et al. \cite{pkmr}), 
we called the single-particle M1 matrix
element given by the Argonne wavefunction as ``experimental" 
although strictly speaking it cannot be so called. 
In what follows, we will use eqs.(\ref{d:exp}) and (\ref{rdexp})
in calculating the Gamow-Teller matrix element with the Argonne wavefunction.

\subsection{The $pp$ $^1S_0$ wavefunction}
\indent

An accurate description of the $pp$ wavefunction involves certain
subtlety and is interesting on its own right. 
KB (\cite{bahcall94}) made an extensive analysis of 
the $pp$ wave function.  Since our result is a bit
different from theirs, we discuss here our approach to this problem
in some detail.

\subsubsection{Potential}
\indent

We begin by decomposing the full potential
$V_{full}$ into four different components:
the nuclear potential $V_N$,
the pure Coulomb potential $V_C$,
the two-photon-exchange Coulomb potential $V_{\rm C2}$,
and the vacuum-polarization potential $V_{\rm VP}$:
\bea
V_{full} &=& V_{N} +V_C + V_{\rm C2} + V_{\rm VP}\label{fullV}
\eea
where (Carlson et al. \cite{carlson}) 
\be
V_C(r) = \frac{\alpha'}{r}, 
\ \ \
\alpha' \equiv \alpha \,
\frac{1 + 2 p^2/m_p^2}{\sqrt{1 + p^2/m_p^2}},
\label{VCdef}\ee
\bea
V_{\rm C2}(r) &=& - \frac{1}{2 m_p^2} \left[
 (\nabla^2 + p^2) \frac{\alpha}{r} +
 \frac{\alpha}{r}  (\nabla^2 + p^2) \right]
\nonumber \\
&=& - \frac{\alpha}{m_p r} V_{full}(r),
\label{C2def} \\
V_{\rm VP}(r)&=& \frac{2\alpha\alpha'}{3\pi r} 
\, \int_1^\infty\! dx\, \e^{-2 m_e r x}
 \left(1 + \frac{1}{2 x^2}\right) \frac{\sqrt{x^2-1}}{x^2},
\label{VPdef}\eea
where, in arriving at the second line of eq.(\ref{C2def}), 
we have removed
the non-local piece $(\nabla^2 + p^2)$
using the equation of motion
\be
(\nabla^2 + p^2) \psi_{pp}(\vec r) = m_p V_{full}(r) \psi_{pp}(\vec r).
\ee
The nuclear potential $V_N$ contains not only strong interactions
but also other electromagnetic interactions such that
the Argonne potential for the $pp$ channel is of the same form as
eq.(\ref{fullV}).\footnote{
$V_{\rm C2}$ and $V_{\rm VP}$ in~(Wiringa et al. \cite{v18})
contain the ``form factor" $F_C (r)$.
We leave out this form factor 
since in our treatment we encounter 
no singularity associated with these potentials.
This difference causes little difference in numerical results.}
In what follows, when we refer to the
``Argonne potential" for this channel, we mean eq.(\ref{fullV}).
\subsubsection{Solution}
\indent

Our treatment of the potential $V_>\equiv V_{\rm C2}+V_{\rm VP}$,
which is somewhat different from that in KB (\cite{bahcall94}),
goes as follows.  First we solve exactly 
the Schr\"odinger equation with the Coulomb plus nuclear potential
\be
V_{\rm C+N}(r)\equiv V_C(r) + V_N(r) 
 = \frac{\alpha}{r} + V_N(r)\label{CN}
\ee
where we have set $\alpha^\prime=\alpha$ 
since we are confining ourselves
to the case $p\approx 0$. 
Then $V_>$, which is of order of $\calO (\alpha^2)$,
is treated perturbatively.

Focusing on the $S$-wave, let $\uC$ be a solution of the
radial Schr\"odinger equation for the potential eq.(\ref{CN})
satisfying the boundary condition $\uC (0)=0$. 
Thus
\be
\left[ \frac{d^2}{dr^2} + p^2 - m_p V_{\rm{C+N}} (r)\right]\uC(r) = 0.
\label{eq:uCNradial}
\ee
Let $r_1\sim (10\sim 20)\ \mfm$ be the range of the nuclear potential
so that
\be
V_N (r)=0, \ \ \ \ r>r_1.
\ee
For $r >r_1$, the solution $\uC$ is given by a linear
combination of the regular and irregular Coulomb wavefunctions,
$F_0$ and $G_0$:
\bea
\uC(r) &=& \phi_{C}(r),\ \ \ r > r_1,
\nonumber \\
\phi_{C}(r) &=& C_0(\eta) \left[G_0(r) + \cot\delta^C\,F_0(r)\right]
\label{asymptot}
\eea
with $C_0(\eta)=\frac{2\pi\eta}{e^{2\pi\eta}-1}$,
$\eta = \frac{m_p \alpha}{2 p}$. Equation
(\ref{asymptot}) defines the normalization of the wavefunction
and the phase shift $\delta^C$.
The properties of the Coulomb functions imply
\bea
\phi_{C}(0) &=& 1,
\nonumber \\
\lim_{r \rightarrow \infty} \phi_{C}(r) &=&
\frac{C_0(\eta)}{\sin\delta^C}
 \sin\left[pr - \eta \ln(2pr) + \sigma_C + \delta^C\right]
\eea
where $\sigma_C \equiv \mbox{Arg}\Gamma(1 + i \eta)$ is the Coulomb phase. 
The scattering length $a^C$
is defined by
\be
\lim_{p\rightarrow 0} C_0^2(\eta) p \cot\delta^C = - \frac{1}{a^C},
\label{aCdef}\ee
with which the asymptotic wavefunction at threshold reads
\bea
 \lim_{p \rightarrow 0} \phi_{C}(r) =
\phi_1(r) - \frac{R}{a^{C}} \phi_2(r)
\label{phipp0}\eea
with
\bea
\phi_1(r) &\equiv&
 \lim_{p\rightarrow 0} C_0(\eta) G_0(r)
 = 2 \sqrt{\frac{r}{R}} K_1(2\sqrt{\frac{r}{R}}),
\nonumber \\
\phi_2(r) &\equiv&
 \lim_{p\rightarrow 0}
  \frac{F_0(r)}{C_0(\eta) p R }
 = \sqrt{\frac{r}{R}}I_1(2\sqrt{\frac{r}{R}}),
\label{phi12}\eea
where $R\equiv (m_p \alpha)^{-1} = 28.8198$ fm 
is the so-called proton Bohr radius, 
and $I_\nu(z)$ and $K_\nu(z)$ are the
modified Bessel functions of order $\nu$. The scattering length
$a^C$ can be determined by matching the logarithmic derivative at
$r = r_1$,
\be
\frac{R}{a^{C}} =
 \left[ \frac{\uC'(r) \phi_1(r)- \uC(r) \phi_1'(r)}{
  \uC'(r) \phi_2(r) - \uC(r) \phi_2'(r)} \right]_{r = r_1} ,
\label{Rapp}\ee
and the effective range $r^{C}$ is given by 
the standard effective range formula
\be
r^{C} = 2 \int_0^\infty\! dr\, \left( \phi_{C}^2(r) - \uC^2(r)
\right).
\label{rCdef}\ee

For the Argonne nuclear potential,
the scattering length and the effective range come out to be
\bea
a^C= -7.8202\ \mfm, \ \ \
r^C= 2.782\ \mfm
\eea
in very good agreement with the ``experimental values"
\bea
a^C_{\rm exp}= -7.8196(26)\ \mfm, \ \ \
r^C_{\rm exp}= 2.790(14)\ \mfm,
\label{arCexp}\eea
given by the Nijmegen multi-energy analysis
(Bergervoet et al. \cite{BCSS}). 
\vskip 0.2cm

In taking into account the effects of the residual long-range
potential, $V_>= V_{\rm C2} + V_{\rm VP}$,
there appear some subtleties.
First, although the magnitudes of $V_{\rm C2}$ and $V_{\rm VP}$
are small [$\sim\calO (\alpha^2)$],
they are both extremely long-ranged,
making it difficult to solve the Schr\"odinger equation
directly. We bypass this difficulty by treating them perturbatively.
Second, depending on how the potential $V_>$ is treated,
there exist in the literature (Bergervoet et al. \cite{BCSS}) 
a number of different definitions of phase shifts 
and low-energy scattering parameters extracted therefrom. 
The $\delta^C$ in eq.(\ref{asymptot})
represents the phase shift due to $V_{C+N}$ relative 
to the phase corresponding to the pure Coulomb potential.\footnote{
One may want to use
the phase shift due to the {\em full} potential relative to the
phase of the Coulomb potential, 
but such a phase shift is inadequate for our present purpose
because it results in an absurdly big scattering length
due to the extremely long range of $V_>$.
On the other hand, it is possible to use 
the phase shift due to the full potential relative to the phase 
corresponding to the Coulomb plus $V_>$ potential.
We shall not adopt this definition because it involves
a much more complicated effective range formula.}
Third, we need to keep track of the changes in the large-$r$ behavior
and in the wavefunction normalization factor
which may cause difference in the flux density.
They should be taken into account self-consistently in the definition
of the cross section, as pointed out by KB (\cite{bahcall94}).
We resolve this problem in such a way that
the $V_>$ potential affects the cross section 
{\em only through the reduced matrix element}, i.e.,
by requiring that $u_{pp}(r)$ -- the radial wavefunction
from the {\em full} potential --
have, modulo change in the phase shift, exactly the same
large-$r$ behavior as that of $\uC(r)$:
\be
\lim_{r \rightarrow \infty} u_{pp}(r) = \frac{C_0(\eta)}{\sin\delta^C}
 \sin\left[pr - \eta \ln(2pr) + \sigma_C + \delta^C + \bar\Delta_0\right]
\label{upplarge}
\ee
where $\bar\Delta_0$ is 
the improved Coulomb-Foldy correction (Bergervoet et al. \cite{BCSS}). 
This means that in order to have the correct flux density, 
eq.(\ref{upplarge}) requires the full $pp$ wavefunction to be of the form,
\bea
\psi_{pp}(r)  &=& \frac{N_{pp}}{r} u_{pp}(r) | {}^1S_0\rangle,
\nonumber \\
N_{pp} &\equiv& \frac{\sin\delta^C}{p C_0(\eta)}
\stackrel{p\rightarrow 0}{=} -C_0(\eta) a^C .
\label{Nppdef}\eea

The $u_{pp}(r)$
that satisfies the boundary condition eq.(\ref{upplarge}) is
\bea
u_{pp}(r) &=& \cos\bar\Delta_0 \, \uC(r) -
 \frac{\sin\delta^C}{p C_0^2(\eta)}
 \left[ \uC(r) P_v(r) + \vC(r) P_u(r) \right],
\nonumber \\
P_v(r) &=& m_p\, \int_r^\infty\! dr' \, \vC(r') V_>(r') u_{pp}(r'),
\nonumber \\
P_u(r) &=& m_p\, \int_0^r\! dr' \, \uC(r') V_>(r') u_{pp}(r')
\label{upp}\eea
where $\vC(r)$ is a second solution of eq.(\ref{eq:uCNradial})
that satisfies the boundary condition
\be
\vC(r) = C_0(\eta)
  \left[\cos \delta^C G_0(r) - \sin\delta^C\,F_0(r)\right], \ \ \ r > r_1 .
\ee
The first order correction in $V_>$ can be obtained by replacing
$u_{pp}$ by $\uC$ in $P_u$ and $P_v$.
The threshold limit is readily obtained
by noticing that $\bar\Delta_0\rightarrow 0$ 
and $\vC(r;\,r>r_1)\rightarrow \phi_1(r)$ when
$p\rightarrow 0$.

A remark is in order here about our treatment of the contribution from the $C2$
potential given in eq.(\ref{C2def}). Up to first
order in $V_{\rm C2}$ and $V_{\rm VP}$, we have
\be
V_{\rm C2}(r) \approx - \frac{\alpha}{m_p r}\left(\frac{\alpha}{r} +
V_N(r)
 \right) \equiv V_{\rm C2}^C(r) + V_{\rm C2}^N(r).
\label{C2prime}\ee
Since the first term, $V_{\rm C2}^C(r)$, is attractive, its
contribution to the phase shift adds destructively to the effect of
the repulsive VP potential. An approximation frequently used in
calculating the phase shifts is to take only this term, ignoring
the second term, $V_{\rm C2}^N(r)$. This is reasonable for scattering
amplitudes since, as one can see from eq.(\ref{upp}), $V_{\rm C2}^N(r)$
cannot affect significantly the large-$r$ behavior of $u_{pp}(r)$.
For the calculation in question, however, 
we need to keep the full potential eq.(\ref{C2prime}). 
The reason is that we
are concerned here with not only the phase shifts but also the
wavefunction in short and intermediate ranges. In fact, we find
that the effect of $V_{\rm C2}^N(r)$ is opposite in sign to, and
greater than, that of $V_{\rm C2}^C(r)$ for $r \lesssim 49\ \mfm$ (except
for very short distances, say, $r\lesssim 0.05$ fm). Consequently,
the C2 contribution to the $pp$ amplitude adds constructively to
the VP contribution. 
These features are exhibited in Fig. 1,
where we plot the ratio $(u_{pp}-\uC)/\uC$
as a function of $r$.

\placefigure{FigWave}  

\section{Numerical Results}\label{numbers}

It is convenient to define the reduced matrix element of
the weak current as
\be
\calM \equiv \calM_{\rm 1B} +\calM_{\rm 2B}
\ee
such that
\bea
\frac{\langle \psi_d|\vec{J}^-|\psi_{pp}\rangle}{\sqrt{8\pi} g_A A_S N_{pp}}
 &=& \left(\hat{s}_d -i\frac{\mu_V}{2g_A m_N}\vec{k}\times \hat{s}_d\right)
     \calM_{\rm 1B} +\hat{s}_d \calM_{\rm 2B}
\nonumber\\ 
&=&\left(\hat{s}_d
-i\frac{\mu_V}{2g_A m_N}\vec{k}\times \hat{s}_d\right) \, \calM
+ \calO \left(Q^4, Q^2|\vec{k}|, |\vec{k}|^2\right),
\label{calMdef}
\eea
where $\hat{s}_d$ is the spin polarization vector of the deuteron
and the normalization factors $A_S$ 
and $N_{pp}$ have been defined in eqs.(\ref{Asdef}) and (\ref{Nppdef}).
Although the reshuffling of terms in the last line 
introduces a term of $\calO (Q^3|\vec{k}|)$ 
(\ie, the product of the WM term and two-body term),
the error made thereby should be negligible.
We give our numerical results in the increasing orders 
in chiral counting.

\subsection{$\calO(1)$ \& $\calO(Q)$ contributions: 
   The single-particle matrix element}
\indent

It follows from eqs.(\ref{1B}) and (\ref{calMdef}) that the
single-particle (reduced) matrix element simplifies to
\bea
{\cal M}_{\rm 1B} &=& \int_0^\infty\! dr\, u_d(r) u_{pp}(r).
\eea
Following the discussion given in the preceding section, we
formally separate the contributions into the 
``Coulomb plus nuclear part" and the ``large-$r$ part"
\bea
{\cal M}_{\rm 1B} &=& {\cal M}_{\rm 1B}^{\rm C+N} + {\cal M}_{\rm 1B}^{>},
\label{onebody} \\
{\cal M}_{\rm 1B}^{\rm C+N} &\equiv& \int_0^\infty\! dr\, u_d(r) \uC(r),
\label{coulomb} \\
{\cal M}_{\rm 1B}^{>} &\equiv& \int_0^\infty\! dr\, u_d(r)
 \left[u_{pp}(r) - \uC(r)\right].\label{residual}
\eea
Clearly the most important term is the ``Coulomb plus nuclear term"
eq.(\ref{coulomb}), so it needs to be calculated as accurately as
possible. This term can in turn be decomposed into a term which can be
extracted from experiments using the  effective range
formula and the remainder that requires theoretical input from
the Argonne potential:
\bea
\calM_{\rm 1B}^{\rm C+N}
= {\cal M}_{\rm 1B}^{{\rm ER}} + \delta {\cal M}_{\rm 1B}^{\rm C+N}.
\eea
If the next-to-leading-order approximation is made to the 
effective range formula, the resulting matrix element is
\bea
{\cal M}_{\rm 1B}^{{\rm ER}} &\equiv&
\int_0^\infty\! dr\, \phi_d(r) \phi_{C}(r)
\nonumber \\
&&+\  \frac12 \int_0^{\infty}\! dr\,
\left\{
\left[\frac{u_d^2(r) + w_d^2(r)}{1+\eta_d^2} -  \phi_d^2(r) \right]
+ \left[\uC^2(r) -  \phi_{C}^2(r)\right] \right\}
\label{INT} \\
&=& 
- \frac{\e^{\zeta}}{a^C \gamma^2}
 \left[1 + \frac{a^C}{R} \left(
 E_1(\zeta) - \frac{\e^{-\zeta}}{\zeta}\right)\right]
-\frac{\rho_d + r^C}{4},
\label{calMER}\eea
where $\zeta \equiv (\gamma R)^{-1}$ and
\be
E_1(\zeta)
 = \int_1^\infty\! dt\,\frac{\e^{-\zeta t}}{t}
 = \int_\zeta^\infty\! dt\,\frac{\e^{-t}}{t}.
\ee
Now we can evaluate ${\cal M}_{\rm 1B}^{{\rm ER}}$ with the
experimental values for the low-energy constants given in
eqs.(\ref{d:exp}), (\ref{rdexp}) and (\ref{arCexp}).
The result is
\be
\left. {\cal M}_{\rm 1B}^{{\rm ER}} \right|_{\rm exp}
 = 5.986(1)\ \mfm - 1.138(5)\ \mfm
\label{MGTER}
\ee
where we have given the individual contributions 
of the two terms in eq.(\ref{calMER}).
If one calculates with the Argonne potential instead of using the
experimental data, one obtains
\be
\left. {\cal M}_{\rm 1B}^{{\rm ER}} \right|_{Argonne}
 = 5.986\ \mfm - 1.137\ \mfm
\ee
which shows once more that it is reasonable to identify the results
given by the Argonne potential with ``experiments".
We can now estimate accurately the remaining small contribution
coming from the difference between eqs.(\ref{coulomb}) and (\ref{INT})
using
the Argonne wavefunctions:
\be
\delta {\cal M}_{\rm 1B}^{\rm C+N}
= 0.011\ \mfm.
\label{MGTremain}\ee
Since this quantity is very  small and the Argonne wavefunction 
is accurate enough,
we have not included the error associated with this correction.
Combining eqs.(\ref{MGTER}) and (\ref{MGTremain}),
we obtain
\be
{\cal M}_{\rm 1B}^{\rm C+N} =
\left(1 \mp 0.02\ \% \mp 0.07\ \%
 \mp 0.02\ \%\right) \cdot
 4.859\ \mfm,
\label{MGTresult}\ee
where the first error comes from the uncertainty in $a^C$,
the second from $r^C$ and the third from $\rho_d$.
(The use of ``$\mp$" instead of ``$\pm$" indicates
that ${\cal M}_{\rm 1B}^{\rm C+N}$ decreases
as each of $|a^C|$, $r^C$ and $\rho_d$ increases.)

Finally the effects of the residual C2 and VP potentials are calculated
from eq.(\ref{residual}):
\bea
{\calM}_{\rm 1B}^> =
  {\cal M}_{\rm 1B}^{\rm VP} +{\calM}_{\rm 1B}^{\rm C2}=-0.031\ \mbox{fm}
\eea
with
${\cal M}_{\rm 1B}^{\rm VP} = -0.022\ \mfm$ and
${\cal M}_{\rm 1B}^{\rm C2} = -0.009\ \mfm$.
For later purpose, we evaluate the ratio 
\bea
\delta_{\rm 1B}
&\equiv& \frac{{\cal M}_{\rm 1B}^{>}}{ {\cal M}_{\rm 1B}^{\rm C+N} }
= \delta_{\rm 1B}^{\rm VP} + \delta_{\rm 1B}^{\rm C2}
 = - 0.63\ \%,
\eea
with
\bea
\delta_{\rm 1B}^{\rm VP} &=& -0.45\ \%,
\nonumber \\
\delta_{\rm 1B}^{\rm C2} &=& 
\delta_{\rm 1B}^{\rm C2:C} + \delta_{\rm 1B}^{\rm C2:N} = 
(0.03\ \%) + (-0.21\ \%)
= -0.18 \ \%,
\label{d1Bv18}\eea
where we have presented each contribution from
$V_{\rm C2}^C$ and $V_{\rm C2}^N$.
It is worth noting that 
the contribution from the C2 potential is about
40 \% of the VP contribution, and 
has the same sign. This is contrary
to what one would
expect if the $V_{\rm C2}^N$ were ignored.

We follow KB (\cite{bahcall94}) and express the result obtained up to this
point in terms of $\Lambda$ defined by
\be
\Lambda \equiv \left(\frac{(a^C)^2\gamma^3}{2}\right)^{\frac12}
 A_S\, \calM_{\rm 1B}.
\label{Lambdadef}\ee
In terms of $\Lambda$,
our result can be summarized as [see eq.(\ref{MGTresult})]\footnote{
If one calculates with the Argonne wavefunctions instead of using the 
experimental values, one finds
$\Lambda^2|_{Argonne} =  (1 + \delta_{\rm 1B})^2\,\times 7.03$.}
\bea
\Lambda^2 &=& (1 + \delta_{\rm 1B})^2\,
\left(1 \pm 0.01\,\%  \mp 0.07\% 
 \pm 0.07\% \right)^2 \times 7.02,\\
&=& (1\pm 0.01\,\% \mp 0.07\,\% \pm 0.07\,\%)^2 \times 6.93
\label{Lam2v18}\eea
where again the errors correspond to the uncertainties
in $|a^C|$, $r^C$ and $\rho_d$ (or $A_S$) in this order.
The only possible sources of theoretical error,
which are due to uncertainties in the Argonne potential,
are in $\delta {\cal M}_{\rm 1B}^{\rm C+N}$ (eq.[\ref{MGTremain}])
and $\delta_{\rm 1B}^{{\rm C2:N}}$ (eq.[\ref{d1Bv18}]). 
If we were to assign conservatively
a $\sim 50 \%$ error to the two quantities there, the resulting
theoretical error on the reduced matrix element would then
be $\sim 0.1\ \%$.

\subsection{Calculation of $\calM_{\rm 1B}^{\rm C+N}$ in effective field theory}
\label{cutoff}
\indent

In Section \ref{2}, we argued 
that the calculation of the single-particle  matrix element 
of the electroweak current using the Argonne wavefunctions is
equivalent to a ``first-principle" calculation using effective field 
theory of QCD. 
How this comes out in the case of the $np$ capture was discussed
in (Park et al. \cite{pkmr}). 
Here we briefly report the result of a first-principle 
calculation for the leading Gamow-Teller matrix element eq.(\ref{coulomb}). 
Such a calculation involving the Coulomb potential
is interesting on its own merit, so we shall describe it
in detail elsewhere (Park et al. \cite{pkmrEFT}). 

As in (Park et al. \cite{pkmr}), 
we shall integrate out all fields other than the matter
field for the nucleon and treat to next-to-leading order (NLO). The potential
has then the contact interactions used for the $np$ channel plus the Coulomb 
interaction
\bea
V(\vec q)&=&
\frac{4\pi}{m_N}\left[C_0 + (C_2\delta^{ij}+D_2\sigma^{ij})q^iq^j\right]
+Z_1 Z_2 \frac{\alpha}{{\vec q}^2},
\nonumber \\
\sigma^{ij} &=& \frac{3}{\sqrt{8}} \left(
\frac{\sigma_1^i \sigma_2^j + \sigma_1^j \sigma_2^i}{2}
- \frac{\delta^{ij}}{3} \vec\sigma_1 \cdot \vec\sigma_2 \right).
\eea
In our case, of course, $Z_1Z_2=1$ for the $pp$ channel
and $0$ for the $np$ channel.
Since the Coulomb interaction is not separable, 
we use the Gaussian regulator as
\bea
V(\vec r)=\int\! \frac{d^3 \vec q}{(2\pi)^3} \,
 S_\Lambda ({\vec q}^{\,2})\, \e^{i{\vec q}\cdot {\vec r}}\, V(\vec q)
\eea
with
\bea
S_\Lambda (\vec q^2)={\rm exp}\left(-\frac{{\vec q}^2}{2\Lambda^2}\right)
\eea
where $\Lambda$ is the cut-off.
The coefficients $C_{0,2}$ for the spin singlet and triplet channels 
and $D_2$ are fixed by $np$ scattering 
and the properties of the deuteron.
The cut-off $\Lambda$ defines a regime in which the effective theory is 
applicable. It is not a free parameter. 
By the rule of (cut-off) effective field theory, 
physical quantities should not depend sensitively 
on the exact form of the cut-off nor
on the exact value of $\Lambda$. 
Since the pion is integrated out, the 
cut-off should be $\sim m_\pi$. 
Indeed as we found in (Park et al. \cite{pkmr}), 
all the static properties of the deuteron are well reproduced
for the range of cut-off, $(140 \sim 200)\ \MeV$.\footnote{
\protect The numerical values of $\Lambda$ figuring here 
could very well be slightly different from the value in 
(Park et al. \cite{pkmr}) because of the slightly different regularizations
involved.
The range of $\Lambda$ found there to give 
correct static properties of the deuteron
was about $(160 \sim 220)\ \MeV$.}
For this range of cut-off, the 
one-body matrix element for the $np$ capture comes out to be 
$4.00 \sim 3.96$
which agrees well with the result of the Argonne wavefunction 
(``experiment"), 3.98. 
The same formalism turns out to give
\bea
\calM_{\rm 1B}^{\rm C+N} ({\rm EFT})= 4.89 \sim 4.85.
\eea
As in the case of the $np$ capture, this agrees closely with
what we might refer to
as  the ``experimental" value, $4.86$, 
with an error of about $0.5\ \%$.
This validates our assertion made in Section \ref{2}.\footnote{
While small in magnitude from the view point of
the next-to-leading order EFT, the error bar of about $0.5\ \%$
is bigger than that with the Argonne potential, $\sim 0.1\ \%$.
This difference may be accounted for by the next-order contribution, namely,
the effective volume term, which we have not considered.}
In our approach, the corrections from the $V_>$ potential are found to be
\bea
\delta_{\rm 1B}^{\rm VP}= -0.45\ \%,\ \ \
\delta_{\rm 1B}^{\rm C2:C}= 0.03\ \%,\ \ \
\delta_{\rm 1B}^{\rm C2:N} = - (0.19 \sim 0.15)\ \%. 
\label{d1Beff}\eea
These values validate the former calculation, eq.(\ref{d1Bv18}),
where we used Argonne wavefunction corresponding to the {\it all} order of
contributions.
Both $\delta_{\rm 1B}^{\rm VP}$ and $\delta_{\rm 1B}^{\rm C2:C}$
are found to be extremely insensitive to the cut-off.
Even $\delta_{\rm 1B}^{\rm C2:N}$ turns out to be 
only mildly dependent on $\Lambda$.
This is somewhat surprising, 
for the effect of $V_{\rm C2}^N \propto \frac{1}{r} V_N(r)$
is expected to be amplified at small $r$ by the nuclear potential.

\subsection{$\calO (Q^3)$ contribution: The exchange-current matrix element}
\indent

The matrix element of the two-body current eq.(\ref{vJ2B}) can be 
separated into three pieces,
\be
\calM_{\rm 2B} =
 \calM_{\rm 2B}^{\rm finite} + \calM_{\rm 2B}^{\rm kin}
 + \calM_{\rm 2B}^\delta
\ee
where $\calM_{\rm 2B}^{\rm kin}$ is the contribution from
the ``kinetic term",
$\calM_{\rm 2B}^\delta$ is the zero-ranged contributions,
and $\calM_{\rm 2B}^{\rm finite}$ is the remaining finite-range
contribution.
{}From eqs.(\ref{calMdef}) and (\ref{vJ2B})
\bea
\calM_{\rm 2B}^{\rm finite} &=&
 \frac{2}{3} \left(\hat c_3 + 2 \hat c_4 + \frac12\right)
 \frac{m_\pi^3}{\mN f_\pi^2}
 \int_0^\infty\! dr\, u_d(r) y_0(m_\pi r) u_{pp}(r)
\nonumber \\
 &-& \frac{2\sqrt{2}}{3}
 \left(\hat c_3 - \hat c_4 - \frac14\right)
 \frac{m_\pi^3}{\mN f_\pi^2}
 \int_0^\infty\! dr\, w_d(r) y_2(m_\pi r) u_{pp}(r),
\label{calM2Bmec}
\\
\calM_{\rm 2B}^{\rm kin} &=&
 \frac{m_\pi^3}{3 \mN f_\pi^2}
 \int_0^\infty\! dr\,
 \frac{y_1(m_\pi r)}{2 m_\pi^2 r}
 \left[u_d(r) u_{pp}'(r) - u_d'(r) u_{pp}(r)\right]
\nonumber \\
&-& \frac{\sqrt{2} m_\pi^3}{3 \mN f_\pi^2}
 \int_0^\infty\! dr\,
 \frac{y_1(m_\pi r)}{2 m_\pi^2 r}
 \left[w_d(r) u_{pp}'(r) - w_d'(r) u_{pp}(r)
   - \frac{3}{r} w_d(r) u_{pp}(r)\right],
\label{calM2Bkin}
\\
\calM_{\rm 2B}^\delta
 &=& - 2 \left(\hat d_1 -2 \hat d_2 + \frac13 \hat c_3
 + \frac23 \hat c_4 + \frac16 \right)
\frac{1}{\mN f_\pi^2}
 \int_0^\infty\! dr\, u_d(r)\, 
\frac{\delta(r)}{4\pi r^2} u_{pp}(r).
\label{calM2Bdelta}
\eea

It is straightforward to calculate 
the matrix elements with the Argonne wavefunctions. 
The results are
\bea
\calM_{\rm 2B}^{\rm finite} &=& 0.246\ \mfm,
\label{finite} \\
\calM_{\rm 2B}^{\rm kin} &=& 0.007\ \mfm,
\label{kinetic}\\
\calM_{\rm 2B}^\delta &=& - (\hat d_1 - 2 \hat d_2 + 0.78)\times
 (0.001\ \mfm).\label{delta}
 \eea
There are two important points to remark here. The first is that the 
finite-range term eq.(\ref{finite}) receives a dominant contribution 
($\sim 93\,\%$) from the deuteron $D$ state. 
This means that the matrix element is somewhat sensitive
to the shorter-range part of the current, \ie,  $y_2(m_\pi r)$. 
We will come back to this later.
The second point is that the zero-range contribution eq.(\ref{delta}) is 
strongly suppressed insofar as the magnitudes of 
the (unknown) constants $\hat{d}_{1,2}$ 
are ``natural", that is $\calO (1)$, as 
required by chiral symmetry. Indeed in the way formulated in 
(Park et al. \cite{pmr_PRL}), 
the zero-range operators should be killed by the short-range correlation 
associated with the very short-distance physics (\ie, degree of freedom)
that is ``integrated out" from the low-energy effective chiral Lagrangian.
We will therefore drop the zero-range term eq.(\ref{delta}) and evaluate the
finite range terms eq.(\ref{finite}) and eq.(\ref{kinetic}) with a short-range 
correlation function incorporated as in (Park et al. \cite{pmr_PRL}). 
As a short-range correlation function
we use a radial wavefunction cut-off
of the form $\theta (r-r_c)$ 
with $r_c\sim (2 m_\pi)^{-1}\sim 0.7$ fm.

In Fig. \ref{FigMEC} is shown the ratio
\be
\delta_{\rm 2B} \equiv \frac{\calM_{\rm 2B}}{\calM_{\rm 1B}}
\ee
as a function of $r_c$.
\placefigure{FigMEC}   
Taking $0.55\ \mfm \le r_C \le 0.8\ \mfm$,
our results may be summarized as
\be
\delta_{\rm 2B} = (4.0 \pm 0.5)\ \% .
\label{delta2B}\ee
This is in the range of the values quoted in the literature 
(Blin-Stoyle \& Papageorgiou \cite{bp65}; 
Gari \& Huffman \cite{gh72}; 
Dautry, Rho, \&  Riska \cite{DRR76}; 
Gari \cite{gar78}). 
We also present the results with other types of 
short-range correlation function $\hat g(r)$
that are frequently used in the literature (Towner \cite{ian}) :
\bea
\delta_{\rm 2B} = 4.84\ \%,&& \ \ \    
\hat g(r) = 1 - j_0(q_c r) = 1 - \frac{\sin(q_c r)}{q_c r},
\\
\delta_{\rm 2B} = 4.01\ \%,&& \ \ \    
\hat g(r) = 1 - \e^{- a r^2} (1 - b r^2),
\eea
where $q_c = 3.93\ \mfm^{-1}$, $a= 1.1\ \mfm^{-2}$ and $b= 0.68\ \mfm^{-2}$.

As mentioned, the non-negligible $r_C$ dependence
in the region $r_C \gtrsim (2 m_\pi)^{-1}$ comes from the fact that 
a short-range (or high-order) current and the $D$-state
component of the deuteron wavefunction are involved.
This is due to the
fact that because of the symmetry involved, $\calO (Q)$ and $\calO (Q^2)$ terms
relative to the single-particle operator are suppressed, an aspect closely 
related to the chiral filter mechanism. 
This means that some terms higher order than 
the $\calO (Q^3)$ ones calculated here -- starting at $\calO (Q^4)$ and probing 
shorter length scales --
may not be negligible compared with the terms 
that are retained.\footnote{
For instance, if one adopts a phenomenological approach
and calculates $\rho$-meson-exchange diagrams involving
a $\rho N \Delta$ vertex (with form factors appended
to all the vertices), 
the one-pion-exchange contribution is found to be 
considerably quenched 
(Bargholtz \cite{bar79}; Carlson et al. \cite{carlson}). 
{}From the chiral perturbation point of view, however,
these diagrams come at $\calO (Q^4)$ or higher, and 
therefore they need to be calculated 
together with all other diagrams of the same chiral order.
We will not attempt such a calculation here.}

\subsection{Cross section}
\indent

For comparison with KB (\cite{bahcall94}), we focus on the low-energy 
cross-section factor $S_{pp}(E)$ (Bahcall and May \cite{bahcall69}),  
\be
S_{pp}(E) \equiv \sigma (E) E \e^{2\pi \eta}
\ee
where $\sigma (E)$ is the cross section for the process (\ref{pp}) and 
$\eta= \frac{m_p\alpha}{2 p}$.  Explicitly, it is of the form
\be
S_{pp}(E) = (1 + \delta_{\rm 2B})^2
 \frac{6}{\pi}\, m_p \alpha\, G_V^2 g_A^2 
\frac{\Lambda^2}{\gamma^3}\, m_e^5 f(E_0 + E),
\ee
where $f(E_0 + E)$ with $E_0\equiv 2 m_p - m_d$ is the
phase volume defined in (Bahcall \cite{bahcall66}) with however 
the WM term contribution
as well as the deuteron recoil taken into account.
These two effects turn out to be very small.\footnote{
The WM term increases $f$ by $0.02\ \%$ while the deuteron recoil
correction decreases it by $0.06\ \%$,
so the net effect is a $0.04\ \%$ decrease.}
Setting $E\approx 0$,  we have 
\bea
S_{pp}^{\chi PT}(0) =4.05 \, \left(\frac{1 + \delta_{\rm 2B}}{1.04}\right)^2
 \left(\frac{g_A}{1.2601}\right)^2
 \left(\frac{\Lambda^2}{6.93}\right) 
 \times (10^{-25} \ \mbox{MeV-barn})
\eea
where we have substituted  $G_V= 1.1356\times 10^{-5}\ {\mbox{GeV}}^{-2}$
and the phase volume $f (E_0)=0.1421$.
It is worth noting, in passing, that our result for the above $S$-factor
is a precise prediction based upon the Standard Model, that is, without
neutrino mass (oscillation).
The corresponding expression given by KB (\cite{bahcall94}) is
\bea
S_{pp}^{KB}(0) =3.89 \, \left(\frac{1 + \delta_{\rm 2B}}{1.01}\right)^2
 \left(\frac{g_A}{1.2573}\right)^2
 \left(\frac{\Lambda^2}{6.92}\right) 
 \times (10^{-25} \ \mbox{MeV-barn}).
\eea
By plugging in the numbers obtained above, 
we arrive at
\bea
S_{pp}^{\chi PT}(0)= 4.05 \,\times 10^{-25}
 (1 \pm 0.15\ \% \pm 0.48\ \%)^2\, \mbox{MeV-barn}
\eea
where the first error arises from the one-body matrix element 
and the second from the two-body term 
due to the uncertainty in the short-range correlation.
(The latter does not include the possible contribution of the next order 
(\ie, $\calO (Q^4)$) term; see later.) This is to be compared with that of KB
\be
S_{pp}^{\rm KB}(0) = 3.89\,\times 10^{-25}
(1\pm 1.1\ \%)\, \mbox{MeV-barn}.
\label{bahcallSS}\ee
 
We are now ready to
make a detailed comparison between our result and that of KB:
\begin{enumerate}
\item The weak-interaction constants used by KB are:
 $g_A=1.2575$ and 
$G_V= 1.151 \times 10^{-5}\ \mbox{GeV}^{-2}$ 
(Freedman \cite{freedman})\footnote{
\protect The value of $G_V$ is not given explicitly in KB (\cite{bahcall94}),
but we can reproduce eq.(\ref{bahcallSS})
only with the use of this value of $G_V$,
a value which was also adopted in (Carlson \cite{carlson}).}. 
Ours are $g_A=1.2601$ and $G_V= 1.136 \times 10^{-5}\ \mbox{GeV}^{-2}$. 
\item KB gave $\Lambda^2=6.92$ and $\delta_{\rm 2B}=1\ \%$
while we have obtained $\Lambda^2=6.93$ and $\delta_{\rm 2B}=4\ \%$.
\end{enumerate}
Apart from small differences in the constants used, the major difference is 
in the exchange-current contribution. In the next section, we will discuss 
whether this is a genuine difference and to what extent our calculation will
be affected by higher chiral-order terms.

Although, as noted by KB (\cite{bahcall94}), 
the corrections due to the long-range potential $V_>$ are small,
it may be of some interest to make a few remarks on its role. 
Our result is that the VP potential decreases the $pp$ rate 
by about $0.9 \ \%$ in agreement with 
(Bohannon \& Heller \cite{bohannon}) 
(where the decrease was found to be about $0.8\sim 1.2\ \%$)
as well as with KB (\cite{bahcall94}),
 who found a $0.9 \ \%$ decrease\footnote{
\protect As mentioned, despite the similarity
in the numerical results, our calculational framework
is different from that of KB. 
This difference will be analyzed in detail
elsewhere (Park et al. \cite{pkmrEFT}).}. 
Our result is slightly at variance with that of (Gould \cite{gould}),
wherein with the use of the WKB approximation 
a $1.3 \ \%$ decrease was found.
In addition the C2 potential lowers the $pp$ rate by about $0.4 \ \%$.
In total, therefore, the downward correction due to $V_>$ is somewhat 
larger than that of KB. Since the $\Lambda^2$ is about the same, 
this means that our single-particle matrix element $\calM_{\rm 1B}$ is 
somewhat larger than that of KB.

\section{Discussion}\label{discussion}

Following the procedure of chiral perturbation theory proven to be highly 
successful for the thermal $np$ capture process, we have calculated the $pp$
fusion rate to $\calO (Q^3)$ in chiral counting relative to the leading 
single-particle Gamow-Teller matrix element. 
{\it To the order considered}, 
the error involved in the calculation is small, $\lesssim 1$ \%. 
This result, given a justification from a cut-off effective field theory 
of low-energy QCD as in the case of the $np$ capture (Park et al. \cite{pkmr}), 
supports the canonical value of Bahcall et al. and does not support the 
``relativistic field theory model" result of 
Ivanov et al. (\cite{ivanov}).

The main caveat in this calculation is in the meson-exchange contribution which
comes out to be about 4 \% when calculated to the chiral order $\calO (Q^3)$.
At the next order, $\calO (Q^4)$, loops and higher-order counter terms enter,
so that there is no reason to believe that they are negligible compared with
the $\calO (Q^3)$ tree contributions. 
(For instance, it could be lowered to about $1\sim 2$ \%  instead of
$\sim 4$ \% found here). 
This aspect is different from the case of the $np$ capture where 
the chiral filter mechanism assures the dominance of the
tree-order pion exchange-current contribution. 
Here the absence of the chiral filter phenomenon {\it can} allow 
higher-order (loop) terms to figure equally importantly as the 
tree-order terms. 
{}From this viewpoint it is not surprising
that a model calculation of 
the terms of $\calO (Q^4)$ and higher 
based on the vector-meson exchange 
and form factors (Bargholtz \cite{bar79}) indicates 
that there can be a considerable suppression 
of the tree-order correction. 
Although such a reduction in the exchange-current contribution
seems to go in the right direction 
for beta decays of higher-mass nuclei (Carlson \cite{carlson}),
it is probably unsafe to import the result of 
such model calculations into our work.
To go to $\calO (Q^4)$ or above in our theory 
at which heavy mesons and form factors come in, 
a large number of Feynman graphs
of the same chiral order have to be computed on the same footing 
to assure chiral symmetry, and such a
calculation has not been done yet. 
In the absence of consistent calculations, our
attitude is that we should not attach any error estimates on terms not
accounted for to the chiral order computed. Calculating the higher-order
terms will be left for a future exercise.

\acknowledgments

We are grateful to John Bahcall and Gerry Brown 
for stressing to us the need to revisit this long-standing problem. 
MR is grateful to Chung Wook Kim, 
Director of the Korea Institute for Advanced Study,
for the invitation to the Institute, 
where part of this paper was written.
TSP is indebted to Francisco Fern\'andez
for the hospitality at 
Grupo de F\'\i sica Nuclear, Universidad de Salamanca,
where this work was completed. 
The work of TSP and DPM was supported in part by the
Korea Science and Engineering Foundation through CTP of SNU and in part by
the Korea Ministry of Education under the grant BSRI-97-2418.
The work of KK was supported in part by NSF Grant No. PHYS-9602000.

\newpage
\begin{figure}[tbp]
\centerline{\epsfig{file=pkmr-pp1.eps,width=5.5in}}
\caption[distortion]{\protect \small
The distortion of the $pp$ wavefunction due to large-$r$ EM
residual interactions defined as $(u_{pp}-\uC)/\uC$ is
shown for various components entering into $u_{pp}$. The solid line
denotes the distortion due to the total (VP+C2) residual
interactions, the dotted line due to the VP, and the dashed line
due to the C2. The contribution from C2 is decomposed into that due
to $V_{\rm C2}^C$ (dot-dashed) and that due to $V_{\rm C2}^N$
(dot-dot-dashed), see eq.(\ref{C2prime}). The C2 contribution
changes from negative to positive at $r \simeq 49\
\mbox{fm}$.}
\label{FigWave}
\end{figure}

\begin{figure}[tbp]
\centerline{\epsfig{file=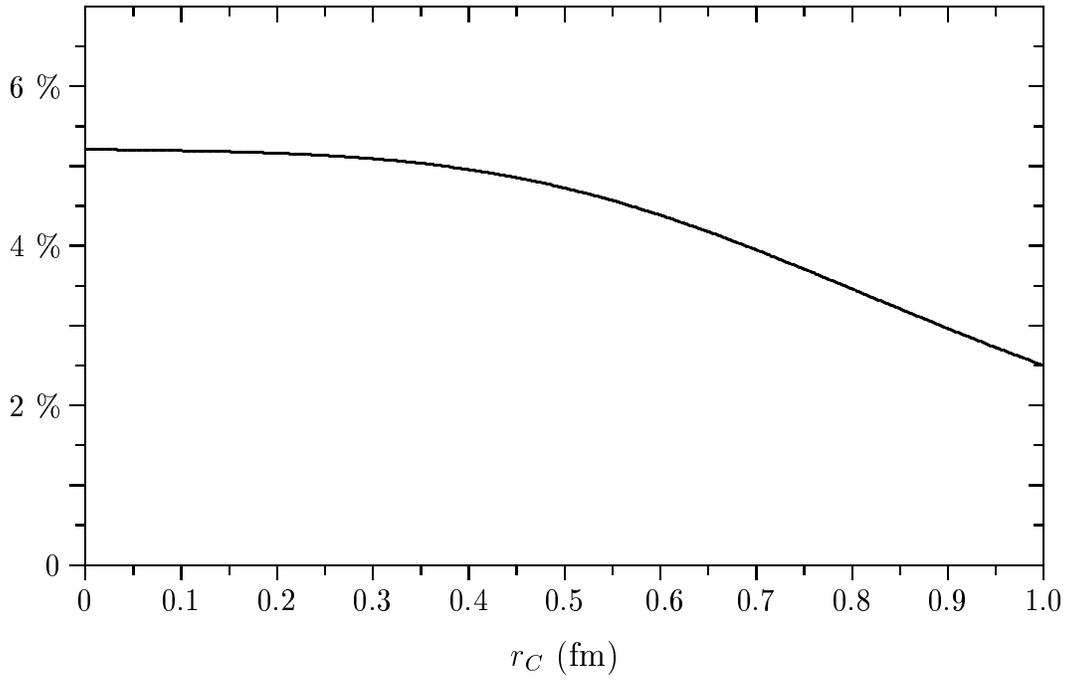,width=5.5in}}
\caption{\protect \small
$\delta_{\rm 2B} \equiv \calM_{\rm 2B}/ \calM_{\rm 1B}$
with varying $r_C$.}
\label{FigMEC}
\end{figure}

\end{document}